\documentclass[12pt,nofootinbib]{revtex4-2}

\usepackage[T1]{fontenc}

\usepackage{color}
\usepackage{colordvi}
\usepackage{amsmath}
\usepackage{amssymb}
\usepackage{amsfonts}
\usepackage{graphicx}
\usepackage{multirow}
\usepackage[dvipsnames]{xcolor}
\usepackage[colorlinks=true,urlcolor=blue,linkcolor=blue,citecolor=blue]{hyperref}
\usepackage{cancel}
\usepackage{booktabs}
\usepackage{caption}
\usepackage{subcaption}
\usepackage{float}
\usepackage{gensymb}
\usepackage{soul}
\usepackage[top=1.135in,bottom=1.17in,left=1.16in,right=1.16in]{geometry}
\numberwithin{equation}{section}

\allowdisplaybreaks

\newcommand{\eq}[1]{\begin{align}#1\end{align}}

\newcommand{\hc}{{\rm h.c.}}

\usepackage[dvipsnames]{xcolor}

\begin{document}
\thispagestyle{empty}

\title{Charged lepton-flavor violating processes and suppression of nonunitary
mixing effects in low-scale seesaw models}

\vspace{50pt}

\author{J. C. Garnica}\email{juangarnica@estudiantes.fisica.unam.mx}\affiliation{Instituto de F\'isica, Universidad Nacional Aut\'onoma de M\'exico, AP 20-364, Ciudad de M\'exico 01000, M\'exico
}\author{G. Hern\'andez-Tom\'e}\email{gerardo.hernandez@cinvestav.mx}\affiliation{Departamento de F\'isica, Centro de Investigaci\'on y de Estudios Avanzados del Instituto Polit\'ecnico Nacional\\
Apartado Postal 14-740, 07000 Ciudad de M\'exico, M\'exico} 
\author{E. Peinado}\email{epeinado@fisica.unam.mx}\affiliation{Instituto de F\'isica, Universidad Nacional Aut\'onoma de M\'exico, AP 20-364, Ciudad de M\'exico 01000, M\'exico
}


\begin{abstract}
We examine the parameter space region of the inverse seesaw model that is consistent with neutrino oscillation data. We focus on the correlation between the current limits from the search of the $\mu\to e\gamma$ lepton flavor violating decay and the non-standard effects associated with the presence of new heavy neutrino states. Unlike what we would expect from an inverse seesaw model, we present a structure for the neutrinos mass matrices in which the rates of charged lepton flavor-violating processes are negligible.  Additionally, we provide a model based on symmetries for such a scenario.

\end{abstract}
\maketitle

\section{Introduction}

The seesaw mechanism offers an attractive scenario to explain the tiny Majorana neutrino masses. The suppression of the light masses is due to the tree-level exchange of very heavy fields, such as right-handed singlet fermions (type-I seesaw)\cite{Minkowski:1977sc, Yanagida:1979as, Gell-Mann:1979vob, Mohapatra:1979ia}, scalar triplet (type-II seesaw)\cite{Magg:1980ut, Schechter:1980gr, Cheng:1980qt, Lazarides:1980nt, Mohapatra:1980yp}, or fermions triplet (type-III seesaw)\cite{Foot:1988aq}. Nevertheless, a direct experimental test of these high-scale scenarios might be impossible due to the decoupling of the new heavy particles.
Alternatively, \emph{low-scale seesaw models}, more specifically, the inverse \cite{Mohapatra:1986bd, Gonzalez-Garcia:1988okv} and linear models\cite{Malinsky:2005bi} are well-motivated variants that open the possibility for a richer phenomenology at a new physics scale accessible to current experiments, such as the existence of new heavy neutrino states with masses at the TeV scale, as well as the presence of charged lepton flavor violating (cLFV) or lepton number violation (LNV) processes at sizeable levels \cite{Ilakovac:1994kj, Deppisch:2004fa, Forero:2011pc, Dev:2012sg, Dias:2012xp, Alonso:2012ji, Abada:2014vea, Hernandez-Tome:2019lkb, Arganda:2014dta, Fernandez-Martinez:2022gsu, Crivellin:2022cve}. 

In this work, we analyzed the parameter space region of the so-called inverse seesaw (ISS) model consistent with the current data in the neutrino sector. We based our analysis on two possible scenarios. In the first one, which we will call \emph{Model A}, we focus on the correlation between non-unitary effects associated with the presence of heavy neutrinos and the limits from the search for cLFV processes.
The second case, \emph{Model B}, presents a  scenario by assuming diagonal structures for the Dirac and heavy mass matrices, while all the structure comes from the lowest scale mass matrix. \emph{Model B} is special since only the SM neutrinos contribute to cLFV processes. Moreover, the contribution from the heavy fermions vanishes at the leading order, which goes against the typical assumption in a low-energy seesaw model. We verified our findings by using two methods, the perturbative block mass matrix diagonalization (BMDM) method presented in \cite{Kanaya:1980cw, Schechter:1981cv} and our complete numerical diagonalization routine implemented in \textit{Wolfram Mathematica}. 

We highlight that when the matrix $\eta$ is diagonal, the cLFV processes are suppressed no matter the seesaw scale. For this reason, we provided a UV-complete model for scenario B. In such a scenario, we stress the comparison of our results using a complete numerical diagonalization with the BMDM.

The structure of this manuscript is as follows: Section \ref{SSM} is devoted to introducing the general form of the mass matrix defining the seesaw models, the basic aspects of the BMDM, and presenting the $\eta$ matrix that quantifies the non-unitary effects. After this, we diagonalised the neutrino mass matrix of the ISS model as a particular case of the general seesaw structure. Section \ref{LFV} presents the analytical expressions, and the current experimental status of the cLFV decays $\ell\to\ell' \gamma$ $\left(\ell=\mu \,(\tau),\, \ell'=e\, (e,\mu)\right)$. In section \ref{NA}, we present the numerical analysis associated with the phenomenology of \emph{Models A} and \emph{B}. Section \ref{Peinado-model} introduces an ultraviolet completion for \emph{Model B}. We finish with a summary and conclusions in Sec. \ref{Conclusions}. 
\section{Seesaw models}\label{SSM}

From a theoretical perspective, the Majorana nature of neutrinos is motivated by the
scale suppression in the dimension 5 Weinberg operator~\cite{Weinberg:1979sa}, whose UV completion may
rise from the seesaw models. In a seesaw model, besides the three active left-handed neutrinos $\nu_{Li}$ ($i=1,2,3$) of the SM, the neutrino sector is extended by a number $n$ of new right-handed singlets fields $N_{Rk}$ ($k=1,2,...,n$) that allow both Dirac and Majorana mass terms:
\eq{
-\mathcal{L}_\textrm{Mass}=\overline{\hat{\nu}_{Li}}\,(\mathcal{M}_D)_{ij}\, \hat{N}_{R_j}+\frac{1}{2}\,\overline{\hat{N}_{Ri}^{c}}\,(\mathcal{M}_R)_{ij}\, \hat{N}_{R_j}+\textrm{h.c.} \,,\label{lgeneralmnu}
}
where $\psi^c \equiv C\overline{\psi}^T$ is the charge conjugated field and $C$ is the charge conjugation matrix. Notice that in such a case, the total number of neutrino states is given by $n'$$\,$$\equiv$$\,$$(3+n)$.

Eq. \eqref{lgeneralmnu} can be written in a compact form, in the basis $\hat{\chi}_L$$\equiv$$\,$$(\hat{\nu}_{L1}, \hat{\nu}_{L2}, \hat{\nu}_{L3}, \hat{N}_{R1}^c, \ldots, \hat{N}_{Rn}^c)$, as follows
\eq{ 
-\mathcal{L}_\textrm{Mass}=\frac{1}{2}\;\overline{\hat{\chi}_L}\; \mathcal{M}\;\hat{\chi}_L^c+\textrm{h.c.} \,, \quad \textrm{where}  \quad
\mathcal{M}_{n' \times n'}=\begin{pmatrix}
0_{3\times 3} & \mathcal{M}_{D_{3\times n}}\\
\mathcal{M}_{D_{n\times 3}}^T & \mathcal{M}_{R_{n\times n}}
\end{pmatrix},\label{GSS}
}
where the hat in Eq. (\ref{GSS}) stands for the fields in the flavor basis, i.e. $\hat{\chi}_{Li}=\mathcal{U}^{\nu^*}_{ij}\chi_{Lj}$, and $\chi_{Lj}$ are the physical neutrino states \footnote{Note also that in order to define all the mass states positive the matrix $\mathcal{U}^\nu$ can be multiplied by a diagonal matrix $\sqrt{\lambda}$ of complex phases, this is equivalent to redefining the fields by $\chi_i\to\chi_{Li}+\lambda_i \chi_{Li}^c$, where $\lambda_i=\pm1$ is the CP parity of the field $\chi_i$. 
}. 
The above structure defines the Type-I seesaw models, where the dimension of the sub-block matrices $\mathcal{M}_D$ and $\mathcal{M}_R$ is denoted by the subindices \footnote{We use this notation in all the text when we consider necessary to clarify the dimensions of the matrices.}, in such a way that the complete neutrino matrix $\mathcal{M}$ has dimensions $n'\times n'$. 
\subsection{Block matrix diagonalization method (BMDM)}\label{BMDM}

In the type-I seesaw, the heavy right-handed neutrinos are integrated out. In this case, we are in the limit $\vert \mathcal{M}_D \vert \, \ll \, \vert \mathcal{M}_R \vert$. It is possible to block-diagonalize the neutrino mass matrix up to terms of the order $\mathcal{M}_R^{-1}\mathcal{M}_D$ by a unitary matrix  $\,\mathcal{U}^\nu$ that connects weak and physical states as follows~\cite{Kanaya:1980cw, Schechter:1981cv}
\eq{
(\mathcal{U}^{{\nu}})^T \mathcal{M}\; \mathcal{U^\nu}=\mathcal{M}^{\textrm{diag}}, \quad \textrm{where}\quad \mathcal{M}^{\textrm{diag}}_{n' \times n'}=\begin{pmatrix}
m_{\nu_{3\times 3}}^\textrm{diag} & 0_{3\times n} \\
0_{n\times 3} & M_{N_{n\times n}}^\textrm{diag}\\
\end{pmatrix}. 
\label{Mdiagonalization} 
}
In the above expression, $m_{\nu_{3\times3}}^\textrm{diag} \equiv \textrm{diag}(m_1,m_2, m_3)$ is a sub-block diagonal matrix associated with the three light active states, while $M_{N_{n\times n}}^\textrm{diag}\equiv \textrm{diag}(m_{N_1},m_{N_2},... m_{N_n})$ is a sub-block diagonal matrix defining the masses of $n$ heavy states. The matrix $\mathcal{U}^\nu$ at leading order is approximated as~\cite{Kanaya:1980cw, Schechter:1981cv} 
\eq{
\mathcal{U}^\nu_{n'\times n'}
&= U^{\nu}_{n'\times n'}\cdot V_{n'\times n'},
\label{U-GSS} 
}with
\eq{
V_{n'\times n'}=& 
\begin{pmatrix}
V_{1_{3\times 3}} & 0\\
0 & V_{2_{n\times n}}
\end{pmatrix},
}
and
\eq{
U^\nu_{n'\times n'}=&\begin{pmatrix}
\mathbb{I}_{3\times 3}-\frac{1}{2}(\mathcal{M}_D^*(\mathcal{M}_R^*)^{-1} \mathcal{M}_R^{-1}\mathcal{M}_D^T)_{3\times 3} & (\mathcal{M}_D^*(\mathcal{M}_R^*)^{-1})_{3\times n}\\
-(\mathcal{M}_R^{-1}\mathcal{M}_D^T)_{n\times 3} & \mathbb{I}_{n\times n}-\frac{1}{2}(\mathcal{M}_R^{-1}\mathcal{M}_D^T \mathcal{M}_D^{*}(\mathcal{M}_R^{*})^{-1})_{n\times n} \\
\end{pmatrix},
}
where $\mathbb{I}$ denotes the identity matrix, while the matrices $V_{1_{3\times 3}}$ and $V_{2_{n\times n}}$ are  unitary matrices connecting the flavor and physical states
\eq{
m_{\nu_{3\times 3}}^{\textrm{diag}}=(V_1^T m_{\nu}V_1)_{3\times 3},\quad\quad
M_{N_{n\times n}}^{\textrm{diag}}=(V_2^T M_{N}V_2)_{n \times n}. 
}
The matrices $m_{\nu_{3\times 3}}$ and $M_{N_{n\times n}}$ are given by
\eq{
m_{\nu_{3\times 3}}=-(\mathcal{M}_{D}\mathcal{M}_{R}^{-1}\mathcal{M}_{D}^T)_{3\times 3},\quad\quad
M_{N_{n\times n}}=\mathcal{M}_{R_{n\times n}}.\label{mnuMN}
}

\subsection{Non-unitarity effects}

The leptonic charged current characterizing a model with three generations of left-handed lepton doublets and $n$ right-handed neutrino singlets can be written as follows~\cite{Schechter:1980gr, Ilakovac:1994kj, Hernandez-Tome:2019lkb}
\eq{
\mathcal{L}_{W^{\mp}}=-\frac{g}{\sqrt{2}}W_\mu^-\sum\limits_{i=1}^3\sum\limits_{j=1}^n B_{ij}\,\overline{\ell_{i}}\,\gamma^\mu\,P_L \chi_{j}+\textrm{h.c.},\label{clc}}
where 
\eq{ \quad B_{ij}=\sum\limits_{k=1}^3 \mathcal{U}^{\ell\hspace{0.015cm} *}_{ki}\,\mathcal{U}^\nu_{kj} \,,\label{clc2}
}
defines the mixing in the leptonic sector, and it has a rectangular form with dimensions $3 \times n'$ ($n'=3+n$ the number of total neutrino states). In eq. (\ref{clc2}), $\mathcal{U}^\ell_{3\times 3}$ is the matrix that diagonalizes the charged lepton mass matrix. It turns out helpful to rewrite the $B_{ij}$ as follows
\eq{
B_{3\times n'}\equiv (B_{L_{3\times 3}},\, B_{H_{3\times n}}),\label{matrixB}
}
where $B_{L_{3\times 3}}$ and $B_{H_{3\times n}}$ are two sub-block matrices describing separately the flavor mixing between light and heavy lepton states, respectively. Therefore, working in the diagonal charged lepton mass basis ($\mathcal{U}^\ell_{ik}=\delta_{ik}$), we have that
\eq{
B_{L_{3\times 3}}&=\left(\mathbb{I}_{3\times 3}-\frac{1}{2}(\mathcal{M}_D^*(\mathcal{M}_R^*)^{-1} \mathcal{M}_R^{-1}\mathcal{M}_D^T)_{3\times 3}\right)\cdot V_{1_{3\times 3}},\label{BL}\\
B_{H_{3\times n}}&=\left(\mathcal{M}_D^*(\mathcal{M}_R^*)^{-1}\right)_{3\times n}\cdot V_{2_{n\times n}}, 
}
with $V_1$ identified with the neutrino mixing matrix $U_{PMNS}$. Furthermore, as with any general matrix, we can write it as the product of a Hermitian matrix and a unitary matrix \cite{Fernandez-Martinez:2007iaa}. $B_{L_{3\times 3}}$ can be rewritten in the following manner 
\eq{
B_{L_{3\times 3}}=(\mathbb{I}_{3\times 3}-\eta_{3\times 3})\cdot V_{1_{3\times 3}}.\label{BL1}
}
In this way, when comparing eqs. \eqref{BL} and \eqref{BL1}, it is clear that the matrix that quantifies the deviation from unitarity of the light neutrino mixing matrix is 
\eq{
\eta_{3\times 3}=\frac{1}{2}\,(\mathcal{M}_D^*(\mathcal{M}_R^*)^{-1} \mathcal{M}_R^{-1}\mathcal{M}_D^T)_{3\times 3}. \label{eta}
}
For the possible phenomenological effects, such as neutrino oscillations, of the non-unitarity neutrino mixing matrix see \cite{Antusch:2006vwa}
\subsection{Inverse seesaw model (\texorpdfstring{$N_R=3, S=3$}{} case)}

A well-motivated variant of the usual (\emph{high-scale}) type-I seesaw is the so-called inverse seesaw (ISS) model~\cite{Mohapatra:1986bd}. In this case, the smallness of the LNV parameter, $\mu$, explains the lightness of neutrinos. This extra suppression of the light neutrino masses allows for heavy neutrino states with masses accessible at current collider energies.

The ISS model requires extending the neutrino sector with right-handed singlet neutrinos $N_{iR}$ and left-handed singlets $S_{j}$. Here, we consider the case with three $N_R$ and three $S_j$ singlets\footnote{Reference ~\cite{Abada:2014vea} studied a minimal scenario with only two $N_{iR}$ and two $S_{j}$ neutrino states.}. The full neutrino mass matrix is~\cite{Forero:2011pc} 
\eq{
\mathcal{M}_{9\times 9}^{\textrm{ISS}}
=\begin{pmatrix}
0_{3\times 3} & M_{D_{3\times 3}} & 0_{3\times 3} \\
(M_D^T)_{3\times 3} & 0_{3\times 3} & M_{3\times 3} \\
0_{3\times 3} & (M^T)_{3\times 3} & \mu_{3\times 3} 
\end{pmatrix},
\label{ISS-mass}
}
with the hierarchy $\vert \mu\vert\ll\vert M_D \vert\ll  \vert M \vert$. We can generalize Eqs. \eqref{mnuMN}, assuming that $M$ is invertible and making the following identification
\eq{
\mathcal{M}_{D_{3\times 6}}^{\textrm{ISS}}= (M_{D_{3\times 3}},0_{3\times 3}),\quad \mathcal{M}_{R_{6 \times 6 }}^{\textrm{ISS}}=\begin{pmatrix}
0_{3\times 3} & M_{3\times 3}  \\
M^T_{3\times 3} & \mu_{3\times 3}  
\end{pmatrix}, 
}
where the inverse matrix of $\mathcal{M}_{R_{6 \times 6 }}^{\textrm{ISS}}$ is given by
\eq{
(\mathcal{M}_R^{\textrm{ISS}})^{-1}_{6\times 6}=\begin{pmatrix}
-((M^T)^{-1}\mu M^{-1})_{3\times 3} & (M^T)^{-1}_{3\times 3}  \\
M^{-1}_{3\times 3} & 0_{3\times 3}  
\end{pmatrix}.
} Using the BMDM for the inverse seesaw model, we have that
\eq{
m_{\nu_{3\times 3}}^{\rm{ISS}}=(M_D (M^{T})^{-1}\mu M^{-1}M_D^T)_{3\times 3}, \quad \textrm{and} \quad M_{N_{6\times 6}}^{\rm{ISS}}=\mathcal{M}_{R_{6\times 6}}^{\rm{ISS}}.
\label{MnuISS}
}
Furthermore, in the limit $\mu \to 0$, the weak charged lepton is given by
\eq{
B^{\rm{ISS}}_{3\times 9}=(B_{L_{3\times 3}}^{\rm{ISS}},\, B_{H_{3\times 6}}^{\rm{ISS}}),
}with
\eq{
B_{L_{3\times 3}}^{\rm{ISS}}=\left(\mathbb{I}_{3\times 3}-\eta^{\rm{ISS}}_{3\times 3}\right)\cdot V_{1_{3\times 3}},\quad  \quad
B_{H_{3\times 6}}^{\rm{ISS}} =
\left(0_{3\times 3},\, (M_D^*(M^{*T})^{-1})_{3\times 3}\right)\cdot V_{2_{6\times 6}}.\label{BHISS} 
}
Whereas the $\eta^{\textrm{ISS}}_{3\times 3}$ matrix is given by
\eq{
\eta^{\rm{ISS}}_{3\times 3}&=\frac{1}{2}(M_D^*(M^{T*})^{-1}M^{-1}M_D^T)_{3\times 3}.\label{eta-ISS}
}
The two cases we will discuss in section \ref{NA} share the assumption that $M$ is diagonal. In such a case, the matrix $V_2$ in Eq. (\ref{BHISS}), required to determine the heavy physical states and their mixings, can be approximated by 
\eq{V_{2_{6\times 6}}=\frac{1}{\sqrt{2}}\begin{pmatrix}
-\mathbb{I}_{3\times 3} & \mathbb{I}_{3\times 3}  \\
\mathbb{I}_{3\times 3} & \mathbb{I}_{3\times 3}  
\end{pmatrix} \begin{pmatrix}
i\cdot \mathbb{I}_{3\times 3} & 0_{3\times 3}  \\
0_{3\times 3} & \mathbb{I}_{3\times 3}  
\end{pmatrix}.}
 The $i$ factor in the last matrix ensures that all masses are positive.
 
\section{Charged Lepton flavor violation processes (cLFV)}\label{LFV}
\subsection{\texorpdfstring{$\ell\to\ell^{'} \gamma$}{} decays}
 
The branching ratio formula of the cLFV processes $\ell\to\ell^{'} \gamma$, with $\ell=\mu \,(\tau),\, \ell'=e\, (e,\mu)$ neglecting the mass of the lighter-charged lepton, is given by 
\cite{Hernandez-Tome:2019lkb}
\eq{
\textrm{BR}(\ell\to\ell'\gamma)&=\frac{\alpha}{\Gamma_{\ell}}m_\ell^3\,\vert F_M^\gamma(0)\vert^2,\\
F_M^\gamma(0)&=\frac{\alpha_W}{16\pi}\frac{m_\ell}{M_W^2}\sum_i B_{\ell i}^*\,B_{\ell'i}\,f_M^\gamma(x_i),\nonumber\\
f_M^\gamma &=\frac{3x^3 \log{x}}{2(x-1)^4}-\frac{2x^3+5x^2-x}{4(x-1)^3}+\frac{5}{6}, 
}where $\alpha=e^2/4\pi$ is the fine structure constant, $\alpha_W\equiv\alpha/s_W^2$, $x_i\equiv m_{\chi_i}^2/M_W^2$ and $m_{\chi_i}$ denotes the mass of all the physical neutrino states.  The current and future limits for these transitions are presented in Table \ref{LIMllpg}.
\begin{table}[]
\begin{center}
\begin{tabular}{ccc}
\hline
Process              & Present Limit        & Future Sensitivity \\ \hline
\hline
$\mu\to e\gamma$     & $4.2\times 10^{-13}$ \cite{
MEG:2013oxv}  & $6\times 10^{-14}$ \cite{MEGII:2018kmf} \\ \hline
$\tau \to e\gamma$   & $3.3\times 10^{-8}$ \cite{BaBar:2009hkt}  & $3\times 10^{-9}$ \cite{Belle-II:2018jsg}  \\ \hline
$\tau \to \mu\gamma$ & $4.2\times 10^{-8}$ \cite{Belle:2021ysv}  & $10^{-9}$ \cite{Belle-II:2018jsg}          \\ \hline
\hline
\end{tabular}
\caption{Present limits and future sensitivities for $\ell\to \ell'\gamma$ decays.}
\label{LIMllpg}
\end{center}
\end{table}
\section{Numerical Analysis}\label{NA}

Let us now discuss the phenomenology of two different scenarios of the ISS model that we call \emph{scenarios A} and \emph{B}. In \emph{scenario A}, the matrices $\mu$ and $M$ in Eq. (\ref{MnuISS}) are diagonals. Therefore all the structure comes from the Dirac mass matrix $M_D$. Notice that to determine which matrices can be considered diagonals by redefining the fields and casting the mass matrices only in terms of physical parameters, we must identify the different transformations that leave invariant the density lagrangian for the leptonic sector, see section 2.4 of reference~\cite{Abada:2014vea}.
 On the other hand, in \emph{scenario B}, we consider that the matrices $M_D$ and $M$ are diagonals, and all the structure comes from the matrix $\mu$. We give an ultraviolet completion for this model in section \ref{Peinado-model}.
\begin{table}[h!]
\begin{center}
\begin{tabular}{c|c|c}
\multicolumn{1}{c}{Parameter} & \multicolumn{1}{c}{Normal ordering (3$\sigma$ range)} & \multicolumn{1}{c}{Inverted ordering (3$\sigma$ range)}\\ \hline \hline
$\sin^2 \theta_{12}$ & 0.271 - 0.369  & 0.271 - 0.369    \\ \hline
$\sin^2 \theta_{23}$ & 0.434 - 0.610  & 0.433 -0.608    \\ \hline
$\sin^2 \theta_{13}$ & 0.02000 - 0.02405 & 0.02018 - 0.02424\\ \hline
$\delta_{CP}/\degree$ &  128 - 359 & 200 - 353          \\ \hline
$\frac{\Delta m_{21}^2}{10^{-5}\textrm{eV}^2}$ & 6.94  - 8.14   & 6.94 -8.14          \\ \hline
$\frac{\vert\Delta m_{31}^2\vert}{10^{-3}\textrm{eV}^2}$ & 2.47 -2.63   & 2.37 - 2.53\\\hline \hline
\end{tabular}
\end{center}
\caption{Neutrino mixing parameters used in our analysis \cite{deSalas:2020pgw}.\label{NOpara} We consider the value $m_{\nu}=0.12/3\, (0.15/3)$ eV as a benchmark for the lightest neutrino mass, taking into account the cosmological limit for the total neutrino mass in the normal (inverted) ordering also reported in \cite{deSalas:2020pgw}.}
\end{table}

\begin{center}{\bf \emph{Scenario A}}\end{center}

The Casas-Ibarra parameterization~\cite{Casas:2001sr} helps to write the Yukawa couplings in terms of the neutrino mass matrix and the other mass matrices in the model as follows   \cite{Deppisch:2004fa, Forero:2011pc} 
\eq{
M_{D_{3\times 3}}=\left(V_1^* \sqrt{m_{\nu}^\textrm{diag}} R^T \left(\sqrt{\mu}\right)^{-1}M ^T\right)_{3\times 3},\label{MD-Casas-Ibarra-par}
}
with $R$ a real $3\times 3$ orthogonal matrix described by three arbitrary rotation angles ($\theta, \phi, \psi$)\footnote{The matrix R is defined, for simplicity, as a $3\times 3$ real orthogonal matrix, similar to reference \cite{Deppisch:2004fa}.}. Moreover, we work on the basis where $M_{3\times3}$ and $\mu_{3\times 3}$ are real diagonal matrices
\eq{
M_{3\times 3}&=\textrm{diag}(M_{11},M_{22}, M_{33})=v_M\cdot\textrm{diag}(1+\epsilon_{M_{11}}, 1+\epsilon_{M_{22}}, 1+\epsilon_{M_{33}})\\
\mu_{3\times 3}&=\textrm{diag}(\mu_{11},\mu_{22}, \mu_{33})=v_\mu\cdot\textrm{diag}(1+\epsilon_{\mu_{11}}, 1+\epsilon_{\mu_{22}}, 1+\epsilon_{\mu_{33}})\label{vmu}.
}
In our analysis, we used sixteen parameters: three mixing angles $\theta_{12},\, \theta_{23},\, \theta_{13}$ and one $CP$-violating phase $\delta_{CP}$, three light neutrino masses in $m_\nu^{\textrm{diag}}=\textrm{diag}(m_{\nu_1},\, m_{\nu_2},\, m_{\nu_3})$, three rotation angles in the $R$ matrix, three parameters for the $\mu$ matrix, and three parameters for the diagonal $M$ matrix. 

We have performed a random scan setting the scale $v_M=1$ (10) TeV, and varying $v_\mu$ into the range $[1, 1000]$ eV, while we choose the rest of the free parameters as follows:
\begin{itemize}
\item  The three light neutrino masses, the three mixing angles, and the Dirac CP-violating phase associated with the active neutrino sector are considered in the range allowed for the current neutrino oscillation data \cite{deSalas:2020pgw} (see Table \ref{NOpara}).
\item The angles $\theta,\,\phi,\, \psi$ in the matrix $R$ vary into the range $[0, 2 \pi]$.
\item The parameters $\epsilon_{M_{ii}}$ $(i=1,2,3.)$ of the matrix $M$ vary into the range [-0.5, 0.5].
\item The parameters $\epsilon_{\mu_{ii}}$ $(i=1,2,3.)$ of the matrix $\mu$ vary into the range   $[-0.5, 0.5]$.
\end{itemize}

\begin{figure}
\begin{center}
\begin{tabular}{cc}
\includegraphics[scale=.5]{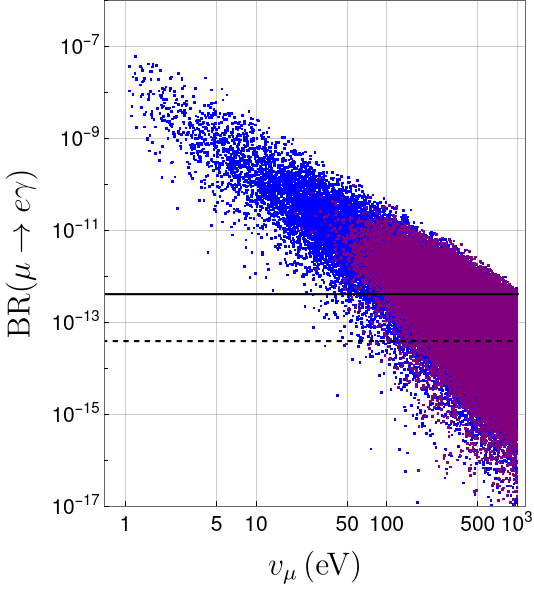} & \includegraphics[scale=.5]{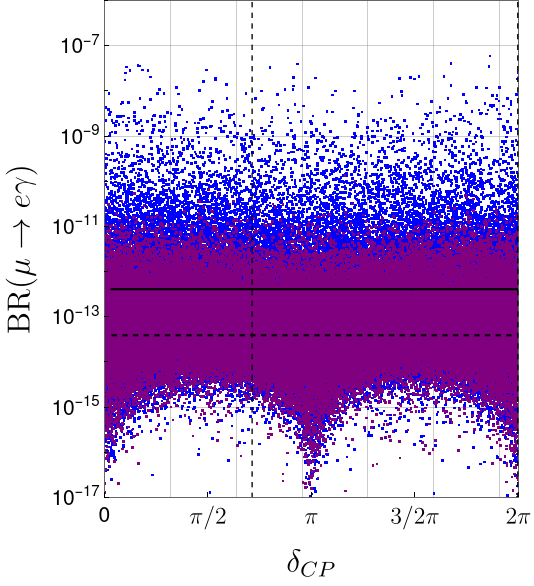}
\end{tabular}
\caption{Branching ratio for the $\mu\to e\gamma$ decay in the inverse seesaw model (\emph{scenario A}). We scan the parameters associated with the neutrino oscillation data assuming the normal hierarchy values shown in Table \ref{NOpara}. Then, we scan the other free parameters as explained in the main text.
The blue (purple) points represent the results setting the scale $v_M=1$ (10) TeV.
The horizontal black solid (dashed) line represents the current limit on BR$(\mu\to e\gamma)<4.2\times 10^{-13}$ \cite{MEG:2013oxv} (future expected sensitivity BR$(\mu\to e\gamma)< 6\times 10^{-14}$  \cite{MEGII:2018kmf}), while the vertical dashed lines in the right plot represent the current limits on $\delta_{CP}$ reported in \cite{deSalas:2020pgw}.}
\label{vmu-d-vsBR}
\end{center}
\end{figure}

At this point, it is worth mentioning that we have done a cross-check of our results using both the BMDM described in section II and our complete numerical routine implemented in Wolfram Mathematica \cite{Mathematica} \footnote{We can share the notebook file with the results upon request.}. Given our scan's numerical matrix in Eq. (\ref{ISS-mass}), we diagonalize it by demanding a high machine precision in extracting its eigenvectors. Then, the matrix $B$ defining the charged lepton current is obtained directly from the Eq. (\ref{clc}), with the sub-block matrices $B_{L_{3\times 3}}$ and $B_{H_{3\times 6}}$ formed by the three first columns, and from the fourth to the nine columns of $B$, respectively. Furthermore, the hermitian $\eta^{\textrm{ISS}}_{3\times 3}$ matrix is obtained directly from the relation
\eq{
B_L \cdot B_L^{\dagger}=\mathbb{I}-2 \eta^{\textrm{ISS}}+\mathcal{O}((\eta^{\textrm{ISS}})^2). 
}
We have considered only points satisfying that $M_{D_{ij}} < 175$ GeV to respect a perturbative limit. After the diagonalization of each numerical matrix in Eq. (\ref{ISS-mass}) of our scan, we observe in the left plot of  Fig. \ref{vmu-d-vsBR} that there are points that easily overpass the current limit set by the MEG collaboration \cite{MEG:2013oxv} for the branching ratio of the $\mu\to e\gamma$ decay \footnote{In these plots, we have found an excellent agreement between the BMDM and our exact numerical method, that is the points of both methods almost overlap. Therefore, we showed only the points obtained with our numerical routine.}. Specifically, setting the scale $v_M \approx 1$ (10) TeV represented by the blue (purple) points, the scale $v_\mu$ must satisfy that $v_\mu \gtrsim 50$ (100) eV to be compatible with both the current limits from $\mu\to e\gamma$ and the data from neutrino oscillation. It is worth noticing that the future sensitivity expected from MEG II will be able to test the points between the solid and dashed black lines in Fig. \ref{vmu-d-vsBR}. 

Additionally, in the right plot of Figure \ref{vmu-d-vsBR}, we show the effects of the $CP$ violating phase $\delta_{CP}$ on the estimation of the $\mu\to e\gamma$ branching ratio. Something interesting to stress here is that more points tend to have a lower decay rate when $CP$ is conserved than when the $CP$ violation is maximal.

\begin{figure}[h!]
\begin{center}
\begin{tabular}{ccc}
\includegraphics[scale=.35]{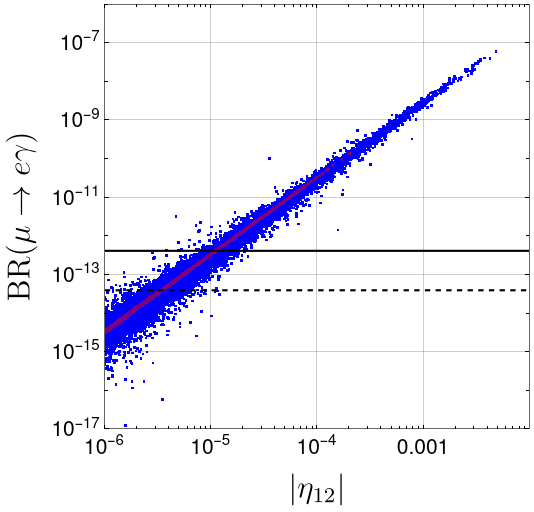} & \includegraphics[scale=.35]{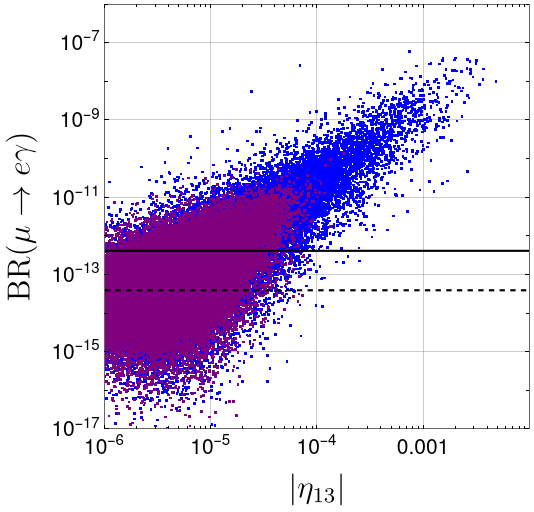}&\includegraphics[scale=.35]{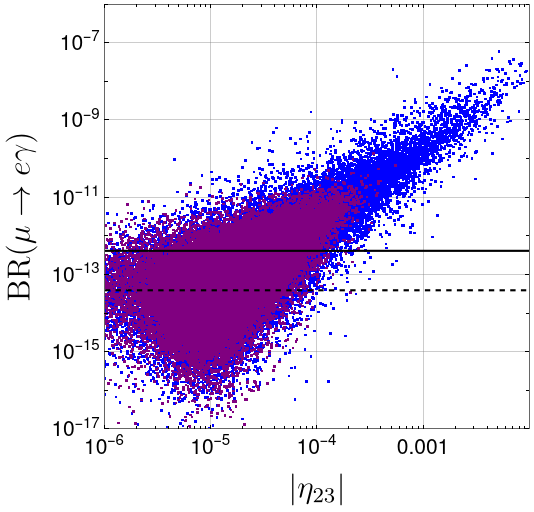}\\
\includegraphics[scale=.35]{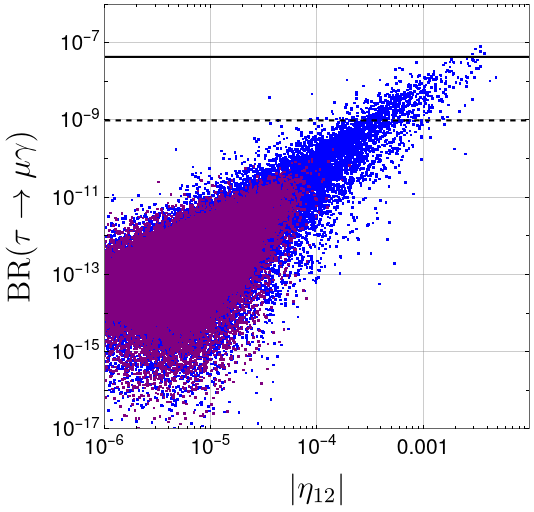} & \includegraphics[scale=.35]{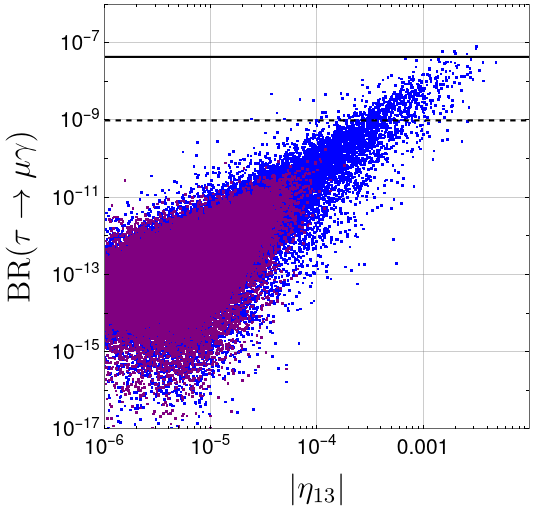}&\includegraphics[scale=.35]{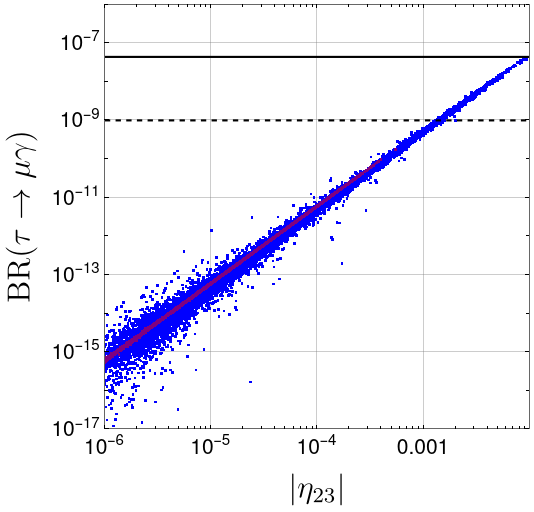}\\
\includegraphics[scale=.35]{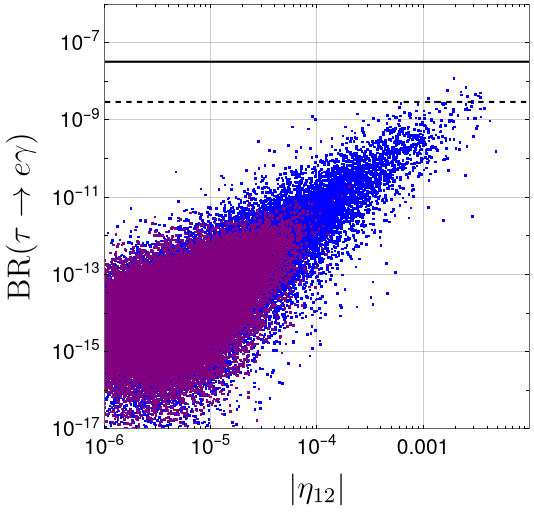} & \includegraphics[scale=.35]{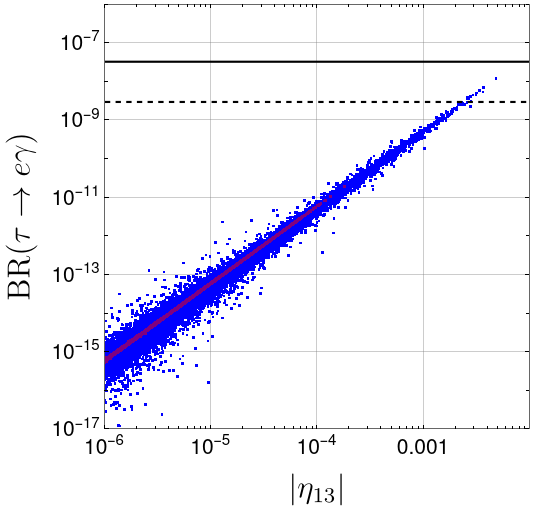}&\includegraphics[scale=.35]{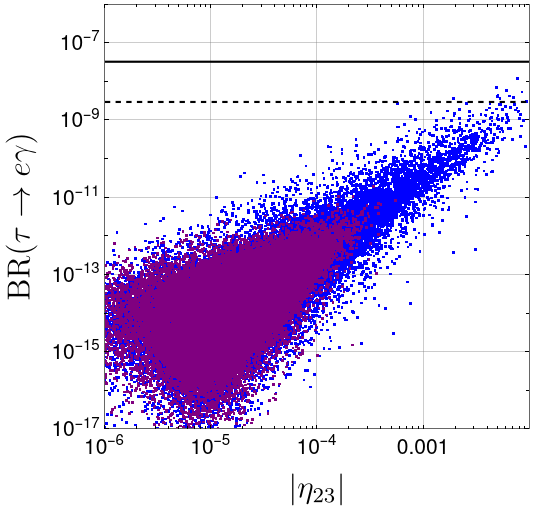}
\end{tabular}
\caption{Branching ratios for the  $\mu\to e\gamma$,  $\tau\to \mu\gamma$, and  $\tau\to e\gamma$ decays in the ISS model (\emph{Model A}) as a function of the absolute value of the elements of the $\eta^{\textrm{ISS}}$. We performed a scan assuming a normal hierarchy for the neutrino oscillation data while we scanned the rest of the free parameters, as explained in the main text. Very similar plots are derived for the inverted hierarchy. Similar to Fig. \ref{vmu-d-vsBR}, the blue (purple) points stand for the results setting the scale $v_M=1$ (10) TeV. The horizontal black solid (dashed) lines represent the current limits (future sensitivities) on BR$(\ell\to \ell'\gamma)$ decays reported in Table \ref{LIMllpg}.
}\label{etavsBR}
\end{center}
\end{figure}

Regarding the correlation between the non-unitary effects and the limits from the search of the $\ell\to\ell'\gamma$  decays, in Fig. \ref{etavsBR}, we show a plot for the branching ratio of the 
$\mu\to e\gamma$ and $\tau\to \ell' \gamma$ ($\ell'=e,\mu$) channels as a function of the absolute value of the elements of the $\eta^{\textrm{ISS}}_{3\times 3}$ matrix. We can see, as expected, that there is a stronger correlation between $\mu\to e\gamma$ and $\eta_{12}$ than with the other elements of the matrix $\eta$. Similarly, with $\tau\to \mu\gamma$ ( $\tau\to e\gamma$) and $\eta_{23}$ ($\eta_{13}$). In fact, according to the current limits taken in our scan and setting the scale $v_M\approx 1$ TeV, we have that the magnitude of non-unitary effects must be $\vert\eta_{12}\vert \lesssim 10^{-5}$,  $\vert\eta_{13}\vert \lesssim 10^{-4}$, and $\vert\eta_{23}\vert \lesssim 10^{-4}$ to respect the most restrictive limit coming from the $\mu\to e\gamma$ channel.

\begin{center}{\bf \emph{Scenario B}}\end{center}

\begin{figure}
\begin{center}
\includegraphics[scale=.6]{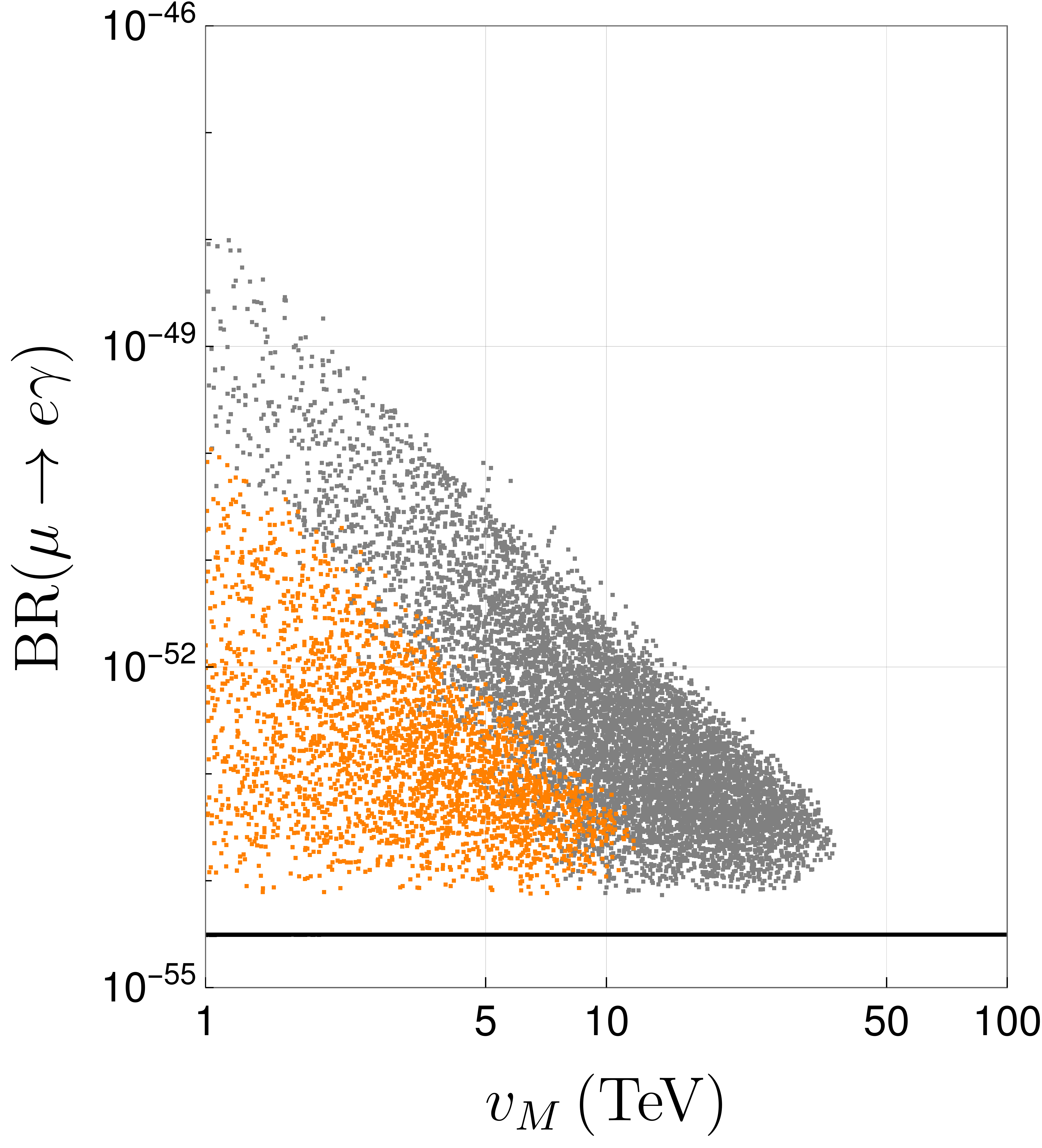}
\caption{Branching ratio for the $\mu\to e\gamma$ decay in the inverse seesaw model (\emph{scenario B}) as a function of the scale $v_M$. We explained the details of our scan in the main text. The black line represents the estimation at leading order in the BMDM, while the orange (grey) points correspond to our complete numerical result assuming the condition $\mu_{ij}<1$ ($10$) MeV.  Namely, using the BMDM approximation for this scenario, only light neutrinos contribute to cLFV processes (black line) because the contribution of the new heavy states is exactly zero in this approximation.
}
\label{vM-vsBRPei}
\end{center}
\end{figure}

Here we consider another case where the Dirac and heavy neutrino mass matrices are diagonal. This is where $M_{3\times3}$ and $M_{D_{3\times 3}}$ in Eq. (\ref{ISS-mass}) are real diagonal matrices
\eq{
M_{3\times 3}&=\textrm{diag}(M_{11},M_{22}, M_{33})=v_M\cdot\textrm{diag}(1+\epsilon_{M_{11}},1+\epsilon_{M_{22}},1+\epsilon_{M_{33}}),\label{vM2}\\
M_{D_{3\times 3}}&=\textrm{diag}(M_{D_{11}},M_{D_{22}}, M_{D_{33}})=\frac{v_{SM}}{\sqrt{2}}\cdot\textrm{diag}(Y_{11}, Y_{22}, Y_{33})\label{vD},
}
where $v_{SM}=246.22$ GeV is the vacuum expectation value of the Higgs field. From the inverse seesaw formula, the $\mu$ matrix is written in terms of $M$, $M_D$ and $m_\nu$ as follows 
\eq{
\mu= M^T M_D^{-1} m_\nu (M_D^{-1})^T M.
\label{mmucase2}
}
In this way, once we give a mass matrix $m_\nu$ compatible with the light neutrino masses and mixings, namely
\eq{
m_\nu= U_{PMNS}^* \textrm{diag}(m_1,m_2, m_3)U_{PMNS}^{\dagger}, 
}
we obtain a valid parameter space for the $\mu$ matrix. In the next section, we give a possible realization of this parametrization based on an Abelian flavor symmetry.

We have also performed a numerical scan for the \emph{scenario B} considering elements of the matrix $\mu_{ij}<1\, (10)$ MeV, grey (orange) points in Figure \ref{vM-vsBRPei} in order to satisfy the condition $\vert \mu \vert \ll \vert M_D \vert$. Here, we let the scale $v_M$ vary from $1$ to $100$ TeV while:
\begin{itemize}
    \item The parameters associated with the neutrino oscillation data, as well as the $\epsilon_{M_{ii}}$ $(i=1,2,3.)$ of the matrix $M$, vary as before.
    \item The Yukawa entries $Y_{{ii}}$ ($i=1,2,3.$) in matrix $M_D$ run into the range [-0.5, 0.5].    
\end{itemize}
   
Something distinctive about this parametrization is that the branching ratio for $\mu\rightarrow e\gamma$ using the leading order BMDM is the same as in the SM. Due to the diagonal structure of the $\eta$ matrix parametrizing the non-unitary effects in Eq. (\ref{eta-ISS}) only the light neutrino states contribute to the cLFV processes \footnote{In this parametrization the $\eta$ matrix is diagonal provided that the product $M_D(M^{T})^{-1}$ is diagonal.}. Indeed, this result is more general, and it will also happen provided that
\eq{
M_D=\lambda M^T,
}
with $\lambda=\textrm{diag}(a,b,c)$ a constant diagonal matrix. On the other hand, we can choose the matrix $\lambda$ to have some off-diagonal non-zero elements to allow specific cLFV processes, see for instance \cite{Arganda:2014dta, Fernandez-Martinez:2022gsu} for a particular example where processes between the two first families ($\mu-e$) are strongly suppressed, but the ($\tau-\mu$) and ($\tau-e$) channels can be maximized within this inverse seesaw parametrization. In Figure \ref{vM-vsBRPei}, we show the estimation for $\mu\to e \gamma$ using both the BMDM and our complete numerical result.  We can see here that using the BMDM, the estimation is of the order $\sim 10^{-54}$ corresponding to the contributions of the light neutrino sector (solid black line). Note, however, that this estimation comes from the assumption of considering the limit when $\mu\to 0$ in the derivation at the leading order of the matrix $\eta^{ISS}$. Contrary, the grey and orange points represent the complete numerical estimation for the points where the elements of the matrix $\mu$ satisfy the condition $\mu_{ij} < 1$ MeV and  $\mu_{ij} < 10$ MeV, respectively.
This plot shows that the results of the exact numerical estimation for the rate of the $\mu\to e\gamma$ can be some orders of magnitude higher for some points in our scan compared with the simple estimation given by the BMDM. In any case, this plot corroborates the fact that, in this parametrization, the contributions of the heavy neutrinos to the branching ratio of the  $\mu\to e\gamma$ decay remain very far from the current and future experimental searches.

\section{Inverse seesaw based on a gauged $U(1)_{B-L}\times Z_5$ symmetry} \label{Peinado-model}

A flavor symmetry can lead to a diagonal Yukawa Lagrangian by appropriately choosing the field representations. For example, in the case of three flavors, the smallest symmetry is $Z_3$, using three different charges, one for each flavor. For the $Z_3$ case, it is sufficient to include a flavon field that transforms non-trivially under the symmetry to generate all entries in the matrix. Therefore, it is possible to fit all light neutrino masses and mixings. 

Let us now consider a generic $Z_N$ and fix $\hat{N}$ according to the case of interest. Using a gauged $U(1)_{B-L}$ symmetry, we will use three RH neutrinos $\hat{N}_R$ with charge $-1$ and three extra sterile fermions $\hat{S}$ with charge $0$, so that they will not contribute to the anomalies.

Regarding $Z_N$, the fermion fields $\hat{L}_i$, $\hat{N}_i$ and $\hat{S}_i$, will transform as $\omega^i$, with $i=1,2,3$. To spontaneously break the $U(1)_{B-L}$ and $Z_N$ symmetries, we need to include two sets of scalar fields $\phi$ and $\xi$\footnote{The number of $\xi$ fields will depend on the specific model.}. We want to reproduce neutrino masses and mixings but also want some correlations between the observables. For this reason, we will use $Z_5$ as flavor symmetry. We show the corresponding $U(1)_{B-L}\times Z_N$ charges for the fields in Table~\ref{UVcompletion-Charges}. In this way, the full Yukawa Lagrangian is
\begin{gather*}
{\cal L}_{Yuk}
=
y^{(\ell)}_i\, \overline{\hat{L}_{i}}\, H\, {\hat{\ell}}_{Ri} + 
y^{(\nu)}_i\, \overline{\hat{L}_{i}}\, \tilde{H}\, \hat{N}_{Ri} +
Y^{(N)}_i\, \overline{\hat{S}_{i}}\, \hat{N}_{Ri}\, \phi + 
 \frac{1}{2} \lambda_{ij}\, \overline{\hat{S}_{i}}\, \hat{S}_{j}^{\,c}\, \xi + \frac{1}{2} \mu_{i,j} \overline{\hat{S}_{i}}\, \hat{S}_{j}^{\,c}+
\hc
\label{lagrangian}
\end{gather*}
where 
$$H=\begin{pmatrix}H^+\\ H^0 \end{pmatrix},\quad L_{i}=\begin{pmatrix}\nu_{Li}\\ \ell_i \end{pmatrix},$$ and $\tilde{H}=i\sigma_2 H^*$.

Here we present a model where the light neutrino phenomenology is compatible with the current experimental data. The mass matrix takes the form of one of the two-zero textures, that is, the $A_1$ in the nomenclature of~\cite{Frampton:2002yf} in which the elements $m_{1,1}$ and $m_{1,2}$ vanish. We obtain the most economical model compatible with this phenomenology by using the symmetry $U(1)_{B-L}\times Z_5$ with the scalar field $\xi$ transforming as $\omega$ under $Z_5$. In this way, the $\mu$ matrix is given by
\begin{equation}
\mu =  \begin{pmatrix}
0 & 0 & \lambda_1 \langle \xi\rangle \\
0 & \lambda_2 \langle \xi\rangle & \mu_{1} \\
\lambda_1  \langle \xi\rangle& \mu_{1} & \lambda_{3} \langle \xi^*\rangle \\
\end{pmatrix}.\label{mumodel}
\end{equation}
This structure in the matrix $\mu$ will lead us to a light neutrino mass matrix with the same structure as Eq. (\ref{mumodel}) corresponding to the $A_1$ two-zero texture in the nomenclature of reference \cite{Frampton:2002yf}. This is compatible with the current experiments on neutrino oscillations for the normal neutrino mass ordering and  predicts negligible neutrinoless double beta decay effective mass parameter since the $m_{\nu_{11}}$ element vanishes at tree-level\cite{Ludl:2011vv,Alcaide:2018vni,DeLaVega:2018bkp}. In this way, we obtained an example of many possible models that one can construct with $M_D$ and $M$ diagonal. Notice that in this model, the $U(1)_{B-L}$ can be local, implying the presence of a new $Z^\prime$ gauge boson.

\begin{table}
\begin{center}
\begin{tabular}{ccc}
  \multicolumn{1}{c}{} &
  \multicolumn{1}{c}{$U(1)_\text{B-L}$} &
  \multicolumn{1}{c}{$\hspace{0.3cm}Z_5\hspace{0.3cm}$}
\\ 
\hline \hline
  \multicolumn{1}{c}{$\hat{L}_{i}=\begin{pmatrix}\hat{\nu}_{Li}\\ \hat{\ell}_{Li}\end{pmatrix}$} &
  \multicolumn{1}{|c|}{$-1$} &
  \multicolumn{1}{c}{$\omega^i$} 
\\ 
\hline
  \multicolumn{1}{c}{$\hat{N}_{Ri}$} &
  \multicolumn{1}{|c|}{$-1$} &
  \multicolumn{1}{c}{$\omega^i$}
\\ 
\hline
  \multicolumn{1}{c}{$\hat{S}_{i}$} &
  \multicolumn{1}{|c|}{$\;\;\,0$} &
  \multicolumn{1}{c}{$\omega^i$}
\\ 
\hline
  \multicolumn{1}{c}{$\phi$} & 
  \multicolumn{1}{|c|}{$-1$} &  
  \multicolumn{1}{c}{$1\;$}
\\ 
\hline
  \multicolumn{1}{c}{$\xi$} & 
  \multicolumn{1}{|c|}{$\;\;\,0$} &  
  \multicolumn{1}{c}{$\omega\;\,$} 
\\ 
\hline \hline
\end{tabular}
\end{center}
\caption{$U(1)_\text{B-L}$$\,$$\times$$\,$$Z_5$ charges for the ISS fields and auxiliary $\phi$, $\xi$ fields.}
\label{UVcompletion-Charges}
\end{table}

\section{Conclusions}\label{Conclusions}

Neutrino oscillations are one of the first pieces of evidence of new physics beyond the original formulation of the Standard Model. Consequently, this evidence raises questions regarding neutrino masses and their Dirac/Majorana nature. There are various massive neutrino models proposed in the literature. Among these, the inverse seesaw model is currently one of the most popular. The idea behind these scenarios is that the physics responsible for neutrino masses could lie on the TeV scale. Such a scenario leads to a testable phenomenology at current or future colliders, for instance, through the search for cLFV processes. 

Our study explores two different scenarios of the ISS model that can accommodate the current neutrino oscillation data but with two entirely different phenomenologies due to the non-unitarity of the light neutrino mixing matrix. In the first case,  where the non-diagonal matrix  $M_D$ is parametrized by Eq. (\ref{MD-Casas-Ibarra-par}), and the assumption that $R$ matrix is real and orthogonal, we have found that cLFV processes take place at sizable levels. Indeed, to be consistent with the limits coming from the current most restrictive $\mu\to e\gamma$ channel, we have found that if the scale of the new heavy states is around $1$ TeV, the magnitude of the non-unitary effects must be $\vert\eta_{12}\vert \lesssim 10^{-5}$.

In the second case, the structure of the model comes from the lowest scale mass matrix $\mu$ described by Eq. (\ref{mmucase2}). We found here that the contributions of the new heavy states to cLFV processes are negligible as a result of the approximate diagonal structure of the matrix describing non-unitary effects.

\section*{Acknowledgements}
This work is supported by the Mexican grants CONACYT CB-2017-2018/A1-S-13051 and DGAPA-PAPIIT IN107621 and IN110622;
We would like to thank Carlos Bunge for helpful discussions about numerical matrix diagonalization methods. 
JCG  is supported by CONACYT. The work of G.H.T. is funded by \emph{Estancias Posdoctorales por M\'exico para la Formaci\'on y Consolidaci\'on de las y los Investigadores por M\'exico, Conahcyt}. EP is grateful for funding from `C\'atedras Marcos Moshinsky' (Fundaci\'on Marcos Moshinsky).




\bibliography{mybib}{}
\bibliographystyle{unsrt}

\end{document}